\documentclass[prd,aps]{revtex4}
\usepackage{amssymb,amsmath,amsfonts,amsbsy,epsfig}
\usepackage{slashed}
\usepackage[dvipsnames]{xcolor}

\begin{document}

\title{Thermodynamics of Graviton Condensate}
\date{\today}

\author{Jorge Alfaro$^{a}$, and Robinson Mancilla$^{a,b}$}

\affiliation{
$^{a}$Instituto de F\'{i}sica, Pontificia Universidad de Cat\'olica de Chile, \mbox{Av. Vicu\~na Mackenna 4860, Santiago, Chile}
}
\affiliation{
$^{b}$Department of Physics, University of California, Santa Barbara, CA 93106, USA}

\begin{abstract} 

In this work, we present the thermodynamic study of a model that considers the black hole as a condensate of gravitons. In this model, the spacetime is not asymptotically flat because of a topological defect that introduces an angle deficit in the spacetime like in Global Monopole solutions. We have obtained a correction to the Hawking temperature plus a negative pressure associated with the black hole of mass $M$. In this way, the graviton condensate, which is assumed to be at the critical point defined by the condition $\mu_{ch}$=0, has well-defined thermodynamic quantities $P$, $V$, $T_{h}$, $S$, and $U$ as any other Bose-Einstein condensate (BEC). In addition, we present a formal equivalence between the Letelier spacetime and the line element that describes the graviton condensate. We also discuss the Kiselev black hole, which can parametrize the most well-known spherically symmetric black holes. Finally, we present a new metric, which we will call the BEC-Kiselev solution, that allows us to extend the graviton condensate to the case of solutions with different matter contents.

\end{abstract}

\maketitle 

\section{Introduction}

In a series of engaging papers, Dvali and Gomez have proposed that black holes (BH) perhaps could be understood as a graviton condensate at the critical point of a quantum phase transition. This idea could explain the thermodynamics properties such as the BH entropy \cite{Bekenstein:1972, Bekenstein:1973, Bekenstein:1975} using tools from condensed matter physics. Somehow, black holes are self-tuned and always stay at the critical point, a distinctive feature of these systems from other quantum systems \cite{Dvali:2012b}. Also, the Hawking radiation \cite{Hawking:1975} would be explained due to the quantum depletion of the gravitons from the condensate. The key idea is that the entire black holes physics can be explained in terms of just one number $N$, the number of "off-shell" gravitons contained in the BEC \cite{Dvali:2011, Dvali:2012a}. This set of ideas are what the authors have called the N-portrait proposal. The authors started to develop this proposal with the strength of graviton-graviton interaction that is measured by a dimensionless coupling $\alpha$ as follows

\begin{equation}\label{interactionGravGrav}
    \alpha\sim\frac{L_{P}^2}{\lambda^2}
\end{equation}

Where $\lambda$ is the characteristic wavelength of the graviton. The Planck length is given by $L_{P}^2=\hbar G$ [$c\equiv 1$]. From this equation, we can see that if the gravitons's wavelength is large, the interaction among gravitons is extremely weak. Consequently,  gravitons behave as free for practical purposes. According to this proposal, the number of gravitons is given by 

\begin{equation}\label{dvaliN}
    N\sim\frac{M r_{g}}{\hbar}
\end{equation}

Where $r_{g}$ is the gravitational radius and $M$ is the total mass of gravitational source. For maximal $N$ the wavelength is such that $r_{g}=\lambda$, then one has that $N\alpha \sim 1$. Black holes must always satisfy this critical condition. They cannot enter into the strong coupling regime [$N\alpha >> 1$]. With the previous equations and some qualitative arguments, Dvali and Gomez achieved to express the Hawking temperature and the BH entropy in terms of the number of gravitons $N$ as follows 

\begin{equation}\label{BECTS}
    T_{h}\sim \frac{1}{\sqrt{N}}, \quad S_{bh}\sim N
\end{equation}

In order to better understand the idea of the graviton condensate, the authors of [8] proposed a more quantitative model of a graviton condensate. In this model, the spacetime is given as background, and over it, a Bose-Einstein graviton condensate is built as a perturbation. In this regard, this model deviates somewhat from the original idea of Dvali and Gomez to accomplish a geometrical approach to the graviton condensate proposal. Besides, this geometrical model was extended in \cite{Alfaro:2019} constructing the graviton condensate for de Sitter spacetime with cosmological horizon and for Reissner-Nordstrom black hole. \\

So far, it has not been computed in the geometrical graviton condensate model an explicit description of the BH thermodynamics, a fundamental aspect to discuss any possible explanation of the ultimate nature of the black holes. Also, if a black hole can be described by a graviton condensate, should not we associate the black hole with a pressure and volume term like any other condensate? if this was the case, one should alter the standard description of BH thermodynamics where there is no associated thermodynamic volume associated with the black hole. The main purpose of this work is to show a consistent thermodynamic description of the graviton condensate using the  Horizon Thermodynamics approach.\\

The paper is organized as follows: in section II, we present the geometrical graviton condensate model briefly, and in section III, we discuss its interpretation proposing some modifications. In section IV, we introduce a local program, called the Horizon Thermodynamics (HT) approach, to the black hole thermodynamics. In section V, we present the thermodynamic study of the graviton condensate and discuss the obtained results. In section VI, we establish a formal equivalence between the line element of the graviton condensate and the Letelier black hole, which describes the Schwarzschild black hole surrounded by a cloud of strings. Section VII is dedicated to introducing the Kiselev black hole, which can parameterize the most well-known spherically symmetric black holes. In section VIII, we introduce a new solution called BEC-Kiselev black hole and study its thermodynamics. Finally, in section IX, we present the conclusions and some remarks.

\section{THE GEOMETRICAL GRAVITON CONDENSATE MODEL}

In this section, we will briefly review the references \cite{Alfaro:2018, Alfaro:2019},  where it is introduced a geometrical model of a graviton condensate.  To achieve this, the metric is split as follows: $g_{\mu\nu}=\tilde{g}_{\mu\nu}+h_{\mu\nu}$. Where $\tilde{g}_{\mu\nu}$ is the background metric and $h_{\mu\nu}$ the quantum fluctuation that describes the graviton condensate. The number of gravitons is proposed to be built from the quantum fluctuation $h_{\mu\nu}$ as follows

\begin{equation}\label{numberPart}
    N=\int^{r_{h}}_{0}\eta dV, \quad \eta\equiv\frac{1}{2r_{h}}h_{\alpha \beta}h^{\alpha \beta}
\end{equation}

Where $\eta$ is the number density of gravitons, $r_{h}$ the event horizon radius and the differential volume is taken as $dV=r^2 \sin(\theta) dr d\theta d\phi$. In order to respect the underlying symmetry of GR, the simplest form to introduce a term related to the BEC, at the level of action, is

\begin{equation}\label{SmatterBEC}
    S_{BEC}=-\frac{1}{8}\int d^4x \sqrt{-\tilde{g}} \mu(x) h_{\alpha \beta}h^{\alpha \beta}
\end{equation}

Where the scalar field $\mu(x)$ would represent the chemical potential of the BEC. The idea behind this construction is that the Gross–Pitaevskii equation, which describes a BEC under certain conditions, is a nonlinear equation such as the Einstein equation. Thus, both equations could be considered as analogous to describe a graviton condensate (see \cite{Alfaro:2018, Alfaro:2019} for the detailed discussion of this proposal). The resulting equation of motion is given by

\begin{equation}\label{originalBECeinsteinEQ}
    G_{\alpha\beta}(\tilde{g}+h)= \Sigma \cdot \mu(x) (h_{\alpha\beta }-h_{\alpha \sigma}h^{\sigma}_{\;\;\beta}), \quad \Sigma\equiv\sqrt{\frac{-\tilde{g}}{-g}}
\end{equation}

We  notice that the RHS of this equation is built from the metric fluctuation, which implies a quite restrictive form for the effective energy-momentum.  Hence, the only mathematically consistent solution for the equation \eqref{originalBECeinsteinEQ} might be $h_{\nu\mu}=0$. However, it was found the following line element

\begin{equation}\label{BECdsNormal}
    ds^2_{BEC}=-\frac{1}{(1-B)}\left(1-\frac{2M}{r}\right) dt^2+\frac{1}{(1-B)\left(1-\frac{2M}{r}\right)}dr^2+r^2d\Omega^2
\end{equation}

Here $B$ is a constant only defined inside the BH. Its magnitude is taken between $0 $ to $1$, otherwise, it could change the sign between $g_{tt}$ and $g_{rr}$. We do not consider the range $[-1,0]$ for $B$ because in this range the local energy density given by $\rho=\frac{B}{8\pi r^2}$ would be negative violating the weak and dominant energy conditions. We call this line element, the BEC-Schwarzschild solution. In addition, we notice that the metric can be written in the following way

\begin{equation}\label{specialEstructure}
    g_{\alpha \beta}=diag\left(\frac{1}{(1-B)}\tilde{g}_{tt}, \frac{1}{(1-B)}\tilde{g}_{rr},\tilde{g}_{\phi\phi},\tilde{g}_{\theta\theta}\right)
\end{equation}

This mathematical structure will be the base to extend this kind of solution to different black holes. The fluctuation part of the metric has a simple mathematical structure

\begin{equation}\label{quantumFluctuation}
    h^{t}_{\;\; t}=h^{r}_{\;\; r}=B
\end{equation}

We provide the result for the factor $\Sigma$ and the scalar field $\mu(x)$, which are 

\begin{equation}\label{chemicalPontential}
    \Sigma=(1-B), \quad \mu=-\frac{1}{(1-B)^2 r^2}
\end{equation}

The nonzero mixed components of the Einstein tensor and the effective energy-momentum tensor are given by

\begin{equation}\label{BECTrr}
        G^{t}_{\;\;t}=G^{r}_{\;\;r}=-\frac{B}{r^2}, \quad T^{t}_{\;\;t}=T^{r}_{\;\;r}=-\frac{B}{8\pi r^2}
\end{equation}

In principle, this solution is valid only inside the event horizon, which is still located at $r_{h}=2M$. The outside solution is the standard Schwarzschild spacetime. Thus, the constant $B$ is zero outside of the event horizon. This setting was chosen because if $B$ is different from zero for the entire spacetime, the metric would not be asymptotically flat. Finally, using equation \eqref{numberPart} we can compute the number of gravitons contained in the BEC

\begin{equation}\label{NforBEC}
    N=\frac{4\pi}{3}4M^2B^2 \Rightarrow M\sim \sqrt{N}
\end{equation}

The constant $ B $ allows us to explicitly connect the number of gravitons with the mass $ M $ of the black hole. Even more remarkable, we can write the Hawking Temperature and the entropy of the black hole in terms of the number of gravitons as follows 

\begin{equation*}
    T_{h}\sim \frac{1}{\sqrt{N}}, \quad S_{bh}\sim N
\end{equation*}

As we mentioned before, the horizon radius $r_{h}$ has not changed, so formally, the entropy of the black hole has not changed either. Therefore, the proportional relation between the number of gravitons $ N $ and the entropy $ S $ is a robust result. On the contrary, it is assumed by hand that Hawking temperature does not change, but the lapse function of the metric has changed with respect to the Schwarzschild metric, so one should expect a different Hawking temperature for this spacetime. \\

\section{INTERPRETATION OF THE MODEL}

So far, we have described the results of \cite{Alfaro:2018, Alfaro:2019} exactly as they were presented by their authors. However, the geometrical graviton condensate model has two interpretation problems: 1) The choice of $B\neq0$ only inside the event horizon introduces a discontinuity in the metric precisely at the event horizon, and 2) The original interpretation considers $\mu(r)$ as chemical potential even when this quantity does not have units of energy. To see the first problem, we express the BEC-Schwarzschild solution in the Eddington-Finkelstein coordinates 

\begin{equation}\label{FEB}
    ds^2_{BEC}=-\frac{1}{1-B}\left(1-\frac{2M}{r}\right)du^2+\frac{2}{1-B}dudr+r^2d\Omega^2
\end{equation}

According to the original interpretation of this model, the constant $B$ is only defined inside the event horizon. However, this setting implies that there is a discontinuity in the metric precisely at the event horizon radius $r_{h}$. We can see this explicitly because $g_{ru}^{BEC}=\frac{1}{1-B}$ inside the horizon, but $g_{ru}^{outside}=1$ outside the horizon, hence $g_{ru}^{outside}(r_{h})\neq g_{ru}^{BEC}(r_{h})$. This discontinuity would imply that radial geodesics would also have a discontinuity which is not acceptable conceptually speaking. Besides, this discontinuity also makes difficult to calculate the Hawking temperature using standard procedures. Then, we must remove this discontinuity by accepting that the parameter $B\neq0$ for the entire spacetime.  Therefore, we must accept that the BEC-Schwarzschild solution is not asymptotically flat. To see the asymptotic nature of the BEC-Schwarzschild metric, we introduce the coordinate transformations $t=\sqrt{1-B}T$ and $r=\sqrt{1-B}R$, then metric \eqref{BECdsNormal} becomes

\begin{equation}
   ds^2_{BEC}=-\left(1-\frac{2M}{\sqrt{1-B}R}\right) dT^2+\frac{1}{\left(1-\frac{2M}{\sqrt{1-B}R}\right)}dR^2+R^2 (1-B) d\Omega^2
\end{equation}

Now, taking the asymptotic limit with $B\neq0$ in the entire spacetime, we obtain 

\begin{equation}\label{Deficit}
   ds^2_{BEC}=-dT^2+dR^2+R^2 (1-B) d\Omega^2
\end{equation}

Therefore, we obtain an angle deficit in the solid angle $\sqrt{1-B}\Omega$. The asymptotic metric (\ref{Deficit}) has the same form of the asymptotic spacetime with Global Monopole \cite{Barriola:1989}, then the parameter $B$ in the graviton condensate spacetime solution is related with a topological defect of the spacetime \cite{Tan:2017}. We recall that Global Monopole spacetime is equivalent to the Letelier spacetime \cite{Letelier:1979}, and in section VI, we will show the equivalence between the BEC-Schwarzschild and the Letelier spacetimes explicitly. Thus, the BEC-Schwarzschild metric is also equivalent to the Global Monopole spacetime, which explains the asymptotic structure of (\ref{Deficit}).\\

Concerning the second interpretation problem, we notice the scalar field $\mu(r)$ apart from not having units of energy also it does not have the appropriate scaling dimension to be consistent with Smarr relation \cite{Smarr:1973, Hawking:1973} which is the integral version of the first law of thermodynamics for black holes. For example, the Hawking temperature scales with the inverse of $r_{h}$ (it is not an intensive thermodynamic variable), and entropy scales with $r^{2}_{h}$, so the product $T_{h} S_{bh}$ must scale with $r_{h}$. (see \cite{Wallace:2018a, Wallace:2018b} for an interesting discussion about these unusual scaling relations of black hole thermodynamics). Now, assume a black hole with chemical potential, then because of the Smarr relation, we know that the product $N\mu_{ch}$ must scale as $r_{h}$, but we also know that $N$ scales with $r_{h}^{2}$ because $S_{bh}\sim N$. Therefore, the chemical potential should scale with $r_{h}^{-1}$ to be compatible with Smarr relations. However, $\mu(r)$ scales with $r^{-2}_{h}$ [see equation \ref{chemicalPontential})]. Therefore, we conclude that $\mu(r)$ cannot be interpreted as the chemical potential.\\

We notice that $\mu(r)$ is a key mathematical ingredient to construct the BEC-Schwarzschild metric, but we also notice that there is no need to interpret $\mu(r)$ as the chemical potential. To understand this, we recall that in the N-portrait proposal of Dvali and Gomez, the black hole is considered to be at the critical point of the quantum phase transition and for large wavelengths, gravitons act as if they were free quasi-particles \cite{Dvali:2011}. Then, we can use the critical condition of a Bose-Einstein condensate made of free particles. So, we recall that the critical temperature of this kind of condensates is defined when the chemical potential vanishes. Therefore, there is no problem that  $\mu(r)$ cannot be considered as the chemical potential because the actual chemical potential vanishes at the critical point of the quantum phase transition. \\

To sum up, in this work, we introduce two re-interpretations concerning original articles of the geometrical graviton condensate: 1) The quantum fluctuation $ h^{t}_{\;\; t}=h^{r}_{\;\; r}=B$ now is defined in the entire spacetime, then this metric is not asymptotic flat. Here, we notice that in the context of Loop Quantum Gravity, a quantum correction of the Schwarzschild black hole has been proposed \cite{Ashtekar:2018}, \cite{Ashtekar:2020}. In those works, the authors also concluded that the quantum fluctuation cannot be turned off at infinity. 2) We consider the graviton condensate as if they were an ideal quantum gas where the critical point for a quantum phase transition is defined when the chemical potential vanishes, so the scalar $\mu(r)$ is not anymore interpreted as the chemical potential. Finally, we notice that if the black hole can be really considered as a graviton condensate, the black hole should have well-defined values for the pressure and the thermodynamic volume as any other condensate. So, with this idea in mind, in the next sections, we will obtain the thermodynamic of the geometrical graviton condensate.\\

\section{HORIZON THERMODYNAMICS APPROACH}

The standard approach of BH thermodynamics is based on integral quantities involved in the Smarr relation, which implies a global notion of spacetime. The Horizon Thermodynamics approach (HT) proposed by Padmanabhan in 2002 is based on local physics \cite{Padmanabhan:2002b}. The key idea of the HT approach is that all horizons must be treated on equal footing because all of them imply a causal disconnection between different patches of the spacetime. Each observer has the right to do physics in her patch, and associate temperature with her notion of horizon \cite{Padmanabhan:2002a}. \\

In the particular case of spherical symmetry, the approach of HT is quite simple. Under this symmetry, the key result is that the thermodynamic pressure of a black hole is identified as $P\equiv T^{r}_{\;\;r}|_{r_{+}}$ \cite{Padmanabhan:2006}, where $r_{+}$ represents the location of the outermost horizon. The first law of HT is 

\begin{equation}\label{FLHT}
    dU=T_{h}dS-PdV
\end{equation}

\begin{equation}
    T_{h}\equiv \frac{\kappa}{2\pi}, \quad S\equiv  \frac{A}{4}
\end{equation}

\begin{equation}\label{TDquantitiesHT}
    V\equiv \frac{4\pi r^{3}_{+}}{3}, \quad  P\equiv T^{r}_{\;\;r}|_{r_{+}}, \quad  U\equiv \frac{r_{+}}{2} 
\end{equation}

The internal energy $U$ is the Misner-Sharp mass evaluated at $r_{+}$. The Misner-Sharp mass is a quasi-local definition of energy that, in spherical symmetry, is well-established and has useful properties \cite{Hayward:1996}. For the Schwarzschild BH, the standard BH thermodynamics approach coincides with the HT approach. The pressure vanishes, and the internal energy is $U=M=E_{k}$,  where $E_{k}$ a is global quantity, the so-called Komar mass. The volume described in equation (\ref{TDquantitiesHT}) must be understood as a thermodynamic volume, not a geometrical volume because this last quantity is not well-defined in general relativity. However, we notice that this thermodynamic volume coincides with the naive expectation for the volume of a spherical black hole.

\section{THERMODYNAMICS OF GRAVITON CONDENSATE}

In the last section, the construction behind of HT approach assumed that $g_{tt}=g_{rr}^{-1}$. However, this is not true in the line element for graviton condensate of equation \eqref{BECdsNormal}. We will derive the HT first law for our case. We write the line element for static spherical symmetry solutions as follows

\begin{equation}\label{ds}
    ds^2=-f(r) dt^2+\frac{1}{h(r)}dr^2+r^2d\Omega^2
\end{equation}

Using the radial component of the Einstein equation, we arrive at 
 
\begin{equation*}
      G^{r}_{\;\;r}=8\pi T^{r}_{\;\;r} \Rightarrow \frac{f(h-1)+rhf'}{r^2f}=8\pi T^{r}_{\;\;r}
      \Rightarrow h-1+\frac{h}{f}rf'=8\pi T^{r}_{\;\;r}r^2
\end{equation*}

We must notice that $\frac{h(r)}{f(r)}=(1-B)^2$. So, it is well-defined to evaluate this fraction at $r=r_{+}$. Taking $T^{r}_{\;\;r}|_{r_{+}}\equiv P$, and recalling that $f(r_{+})=h(r_{+})=0$, we arrive at

\begin{equation*}
     \Rightarrow \frac{(1-B)^2}{2}r_{+}f'(r_{+})-\frac{1}{2}=4\pi P r^{2}_{+}
\end{equation*}

Finally, multiplying the whole equation with $dr_{+}$, and reorganizing the differential in a suggesting way, we have the following expression

\begin{equation*}
  \Rightarrow   (1-B)^2\frac{f'(r_{+})}{4\pi}d(\pi r_{+}^2)-d(r_{+}/2)=P d\left(\frac{4}{3}\pi r_{+}^3\right)
\end{equation*}

In this way, we recover the first law of HT, which is given by

\begin{equation*}
    dU=T_{h}dS-PdV, \quad S\equiv  \frac{A}{4}, \quad V\equiv \frac{4\pi r^{3}_{+}}{3}
\end{equation*}

Where the internal energy and the Hawking temperature are given by

\begin{equation}\label{misnermassBEC}
    U\equiv\frac{r_{+}}{2} \Rightarrow U=M
\end{equation}

\begin{equation}\label{BECtemperature}
    T_{h}\equiv \frac{(1-B)^2 f'(r_{+})}{2\pi} \Rightarrow T_{h}=\frac{(1-B)}{8\pi M }
\end{equation}

We use the equation \eqref{BECTrr} to evaluate the pressure as the HT approach stated to do, which is given by

\begin{equation}\label{BECpressure}
    P\equiv T^{r}_{\;\;r}|_{r_{+}} \Rightarrow P=-\frac{B}{32\pi M^2} 
\end{equation}

We also notice the following interesting relation

\begin{equation}
    P=-\frac{B}{3}\frac{M}{V} 
\end{equation}

This equation looks like the relation between pressure and energy density in the case of electromagnetic radiation. Finally, using equation \eqref{NforBEC} we can express each thermodynamic quantity in terms of $N$ 

\begin{equation}
    M\sim \sqrt{N},\quad  S\sim N,  \quad T_{h}\sim \frac{1}{\sqrt{N}}
\end{equation}

\begin{equation}
   V\sim N^{3/2}, \quad P\sim \frac{1}{N} 
\end{equation}

These relations are entirely satisfactory from the point of view of the N-Portrait proposal \cite{Dvali:2011, Dvali:2012a, Dvali:2012b}. Therefore, the black hole thermodynamics can be parameterized by just one number: the number of gravitons of the BEC given by $ N $. We have successfully associated a volume and a pressure with the graviton condensate.  We can see that $P$ given by equation \eqref{BECpressure} is negative. Then, we could consider this negative pressure as a tension instead of pressure. Perhaps, this negative pressure does not allow the matter to fall inside of the BH. We remind that the BEC-Schwarzschild black hole is a vacuum solution that considers a quantum fluctuation given by $h_{\mu \nu}$; it does not contain any type of matter inside of it. \\

It may be impressive to associate a negative pressure with a black hole without matter content. However, several authors have reached similar conclusions in alternative models to traditional black holes. In 2000, Chaplin and colleagues argued that a black hole would be a quantum phase transition of the vacuum spacetime. According to them, the black holes would have a negative pressure similar to Bose fluid at the critical point \cite{Chaplin:2000}. Another proposal that requires a negative radial pressure is the "gravastars" model developed by Mazur and Mottola \cite{Mazur:2015}. The motivation of this idea is to eliminate the central singularity changing the interior black hole solution by another solution similar to de Sitter spacetime. In a recent paper, Brustein and colleagues have argued that in order to avoid the gravitational collapse, a considerable radial negative pressure is necessary \cite{Brustein:2019}. In the same article, they proposed a black hole model made of closed interacting strings with state equation $\rho=-P_{r}$. There are also other recent proposals with negative pressure associated with black holes, such as \cite{Vaz:2014} and \cite{Leclair:2019}. We can conclude that a common characteristic for many alternative models of black holes is the presence of a negative pressure term. Most of them require negative pressure to avoid the central singularity. Other models relate the negative pressure with some quantum phase transition.\\

Ashtekar and coworkers have obtained a quantum correction for the Schwarzschild black hole in the context of Loop Quantum Gravity, which affects both interior and exterior solution \cite{Ashtekar:2018, Ashtekar:2020}. They computed a quantum correction to the Hawking temperature, which is given by

\begin{equation}
    T_{h}=\frac{1}{8\pi M}\frac{1}{1+e_{M}}, \quad e_{M}=\frac{1}{256}\left(\frac{\gamma \Delta ^{1/2}}{\sqrt{2\pi}M}\right)^{8/3}
\end{equation}

Where $\Delta\approx 5,17 L_{p}^{2}$ is the quantum area gap, and $\gamma\approx0.2375$ is the Barbero-Immirzi parameter of LQG. Then,  $e_{M}$ is only defined in terms of the mass $M$. From the previous expression, we can make the following Taylor expansion 

\begin{equation}
    T_{h}\approx\frac{(1-e_{m})}{8\pi M}
\end{equation}

This expansion is possible because $e_{M}$ is an extremely small value. For a solar mass black hole, $e_{M}\approx 10^{-106}$. This expression is exactly the same that our result for the Hawking temperature given by equation \eqref{BECtemperature} However, they are obtained in a very different way. Even so, we can estimate the possible value of $ B $ assuming that $B\approx e_{M}$. Therefore, for a solar mass black hole, $B$ would be of order $B\approx 10^{-106}$. From this estimation, we can conclude that the parameter $B$ related to the graviton condensate would be naturally suppressed in our universe. Then, $B$ would affect in a negligible way any classical test of GR. \\

It is known that the Misner-Sharp mass is equal to the Komar mass only for the case of Schwarzschild black hole i.e. spacetimes with $T_{\mu\nu}=0$. However, for the BEC-Schwarzschild black hole, we obtained  $U_{BEC}=M$ which looks like a contradictory result because to obtain this solution we have used an effective energy-momentum tensor given by equation (\ref{BECTrr}). To clarify this point, we will compute the Komar mass for the graviton condensate. We start making the following coordinate transformation $t=\frac{T}{1-B}$ in the line element of equation \eqref{BECdsNormal}. Then, we obtain 

\begin{equation}
    ds^2_{BEC}=-(1-B)\left(1-\frac{2M}{r}\right) dT^2+\frac{1}{(1-B)\left(1-\frac{2M}{r}\right)}dr^2+r^2d\Omega^2
\end{equation}

This form allows us to apply straightforwardly the Komar mass formula by identifying the lapse function as

\begin{equation}\label{LapseFunction}
    f(r)=(1-B)\left(1-\frac{2M}{r}\right)
\end{equation}

Then, using the Komar mass formula with $r_{h}=2M$, we obtain that 

\begin{equation}
    E_{k}=\frac{r^{2}_{h}}{2}f'(r_{h}) \Rightarrow E_{k}=M(1-B)
\end{equation}

Therefore, Komar mass for the graviton condensate is different from the Misner-Sharp mass $U=M$ as we could expect from previous considerations.  Both energy quantities satisfy the following relation

\begin{equation}\label{energies}
    U=E_{k}+BM
\end{equation}

What does the extra factor $BM$ mean in equation \eqref{energies}? We propose to compute the following integral 

\begin{equation}
   E^{*}\equiv \int^{r_{h}}_{0}\rho_{local} dV, \quad \rho_{local}=\frac{B}{8\pi r^2}
\end{equation}

Where $\rho_{local}$ is the local energy density that is positive-defined [recall $B\in [0,1)$], and we use the differential "thermodynamic" volume $dV=r^2 \sin(\theta)dr d\theta d\phi$ to obtain that

\begin{equation}
   E^{*}=  \frac{B}{2}r_{h} \Rightarrow E^{*}=BM
\end{equation}

We have obtained the correct term $BM$, which appears in equation \eqref{energies}. The term $E^{*}=BM$ that could be interpreted as energy associated with the quantum fluctuation $h_{\nu\mu}$. We also notice that from the lapse function of equation (\ref{LapseFunction}), we recover the Hawking temperature straightforwardly by using the Hawking's law $T_{h}=\frac{\kappa}{2\pi}$ and $\kappa=\frac{f'(r_{h})}{2}$. Thus, the standard procedures and the Horizon Thermodynamic approach compute the same Hawking temperature as one could expect. Furthermore, we notice that the time-Killing vector $\xi^{\alpha}_{(T)}$ has the following normalization

\begin{equation}\label{Normalization}
   \xi^2_{(T)}=-(1-B)
\end{equation}

As we discussed in section III, this spacetime is not asymptotically flat. There is an angle deficit in the solid angle of  $\sqrt{1-B}\Omega$. Thus, there is not a universal way to define the normalization of the time-killing vector. The particular normalization of equation (\ref{Normalization}) makes it possible to obtain the same Hawking temperature by using the local computation of the Horizon Thermodynamics, by using the Hawking's formula or Euclidean tools, and also by using the results from the thermodynamic quantities of Letelier spacetime that we will show in the following section. As a final comment, there is nothing new in the fact that $\xi^2_{(T)}$ does not have a unit normal. In fact, in the AdS-Schwarzschild black hole the normalization of the time-killing vector blows up in AdS boundary as $\xi^2_{(t)}|^{Schw-AdS}\sim\frac{|\Lambda|}{3}r^2.$

\section{THE FORMAL EQUIVALENCE WITH LETELIER BLACK HOLE}

We will recall the calculation of a black hole surrounded by a cloud of strings based on \cite{Bezerra:2016}. We start considering a moving infinitesimally thin string  that traces out a two-dimensional world sheet $ \Sigma $, which is parameterized as follows

\begin{equation*}
    x^{\mu}=x^{\mu}(\lambda^{a}), \quad a=0,1 
\end{equation*}

Where $\lambda^{0}$ is a time-like parameter and $\lambda^{1}$ a space-like parameter. Assuming the action depends only on $\lambda^{0}$ and $\lambda^{1}$, the Nambu-Goto action is proportional to the area of the worldsheet expanded by the string motion. Therefore, we have

\begin{equation}\label{NambuGotoAction}
    S_{NG}=D\int_{\Sigma}\sqrt{-\gamma}d\lambda^{0} d\lambda^{1} 
\end{equation}

Where $D$ is a positive constant related to the  string tension, and $\gamma$ is the determinant of the induced metric. It is defined the bi-vector $\Sigma^{\mu\nu}$ as follows

\begin{equation*}
    \Sigma^{\mu\nu}\equiv\epsilon^{ab}\frac{dx^{\mu}}{d\lambda^{a}}\frac{dx^{\nu}}{d\lambda^{b}}
\end{equation*}

Using this definition in the Nambu-Goto action, we obtain an alternative form to the action

\begin{equation*}
    S_{NG}=D\int_{\Sigma}\sqrt{-\frac{1}{2}\Sigma_{\mu\nu}\Sigma^{\mu\nu}}d\lambda^{0} d\lambda^{1} 
\end{equation*}

From this action, one can get an effective energy-momentum tensor for a cloud of string, and then, to consider spherical symmetry to solve the Einstein equation. The solution is called the Letelier black hole which has the following line element

\begin{equation}
    ds^2=-\left(1-a-\frac{2m}{R}\right) dT^2+\frac{1}{\left(1-a-\frac{2m}{R}\right)}dR^2+R^2d\Omega^2
\end{equation}

Here $a$ is an adimensional parameter related to the energy density of the cloud of string, hence, it is demanded $a>0$. We notice that in the standard Nambu-Goto action there is an overall negative sign that here is not considered, and there is no problem at all, because the parameter $a$ is a integration constant independent of the value and sign of the constant $D$ (See a detailed discussion in \cite{Bezerra:2018}). According to Letelier's interpretation, this solution represents the Schwarzschild black hole surrounded by a cloud of strings \cite{Letelier:1979}. \\

Supposedly, $m$ represents the mass of the black hole. To arrive at this conclusion, one takes $a=0$ and demands to recover the Schwarzschild solution. This conclusion cannot be obtained using the asymptotically flat limit, because $a$ does not disappear in this limit. Therefore, the Letelier spacetime is not asymptotically flat, which introduces an ambiguity to define $m$ as the mass of the BH. Then, we have the right to define $m=M(1-a)$ and obtain that 

\begin{equation*}
    ds^2=-\left(1-a-\frac{2M(1-a)}{R}\right) dT^2+\frac{1}{\left(1-a-\frac{2M(1-a)}{R}\right)}dR^2+R^2d\Omega^2
\end{equation*}

In this expression, we can still recover the Schwarzschild solution demanding that $a=0$, where obviously $m=M$. Factorizing the term $(1-a)$ in this line element, and introducing the following coordinate transformation $T=\frac{t}{\sqrt{1-a}}$ (and $R=r$), we obtain 

\begin{equation*}
    ds^2=-\frac{1}{(1-a)}\left(1-\frac{2M}{r}\right) dt^2+\frac{1}{(1-a)\left(1-\frac{2M}{r}\right)}dr^2+r^2d\Omega^2
\end{equation*}

This line element is formally the same as that of the BEC-Schwarzschild BH [See equation \eqref{BECdsNormal} with $ a=B $]. Is this a mere coincidence? In the Letelier spacetime, the energy-momentum tensor is due to a cloud of strings with their respective worksheet. In the graviton condensate, the energy-momentum tensor is built from the quantum fluctuation of the background metric given by $h_{\mu\nu}$. Besides, the Letelier solution comes from of a Nambu-Goto action \eqref{NambuGotoAction}. In the graviton condensate, we use the action given by \eqref{SmatterBEC}. They are different actions, even more important: they are entirely different at a conceptual level.  Could it be possible that quantum fluctuations of metric are due to a cloud of strings? We will leave this possibility for later works. Finally, we write down the Hawking temperature and black hole entropy for this Letelier spacetime

\begin{equation}
    T_{h}= \frac{(1-a)^2}{8\pi m}, \quad S=\frac{4\pi m^2}{(1-a)^2}
\end{equation}

These results where obtained in \cite{Bezerra:2018}. If we identify $m=M(1-a)$ as before, we clearly recover the same results for the entropy and Hawking temperature computed in section V for the geometrical graviton condensate. 

\section{KISELEV BLACK HOLE}

The multi-components Kiselev black hole has the following lapse function 

\begin{equation}\label{LapseFKiselev}
    f(r)=1-\frac{2M}{r}-\sum_{i}\frac{C_{i}}{r^{3\omega_{i} +1}}
\end{equation}

The energy-momentum tensor which generates this solution is given by

\begin{equation}\label{EnergyMomentumKiselev}
    T^{t}_{\;\;t}=T^{r}_{\;\;r}=\sum_{i}\rho_{i},\quad T^{\theta}_{\;\;\theta}=T^{\phi}_{\;\;\phi}=-\frac{1}{2}\sum_{i}\rho_{i}(3\omega_{i}+1), \quad \rho_{i}\equiv\frac{3C_{i}\omega_{i} }{8\pi r^{3(\omega_{i}+1)}}
\end{equation}

Kiselev obtained this solution in 2002 \cite{Kiselev:2002}. We call the parameter $\omega_{i}$ the state parameter, and $C_{i}$ the Kiselev charge. According to Kiselev, when the only nonzero state parameter is taken between $\omega_{1}\in (-1,-1/3)$, this solution represents the Schwarzschild black hole surrounded by the quintessence. For the particular case of $\omega_{1}=-\frac{1}{3}$, the lapse function and energy-momentum tensor become

\begin{equation*}
    \Rightarrow f(r)=1-\frac{2M}{2}-C_{(quint)}r
\end{equation*}

\begin{equation*}
    \Rightarrow T^{t}_{\;\;t}=T^{r}_{\;\;r}=2T^{\phi}_{\;\;\phi}=2T^{\theta}_{\;\;\theta}=-\frac{2C_{(quint)} }{8\pi r}
\end{equation*}

The Kiselev BH is used as a toy model to study different aspects of BH. Virtually all publications on this model have preserved the original interpretation. Recently,  Matt Visser pointed out that this interpretation is inadequate \cite{Visser:2019}. The quintessence is a scalar field that has associated a perfect fluid type energy-momentum tensor \cite{Hervik:2010}. However, we can see from equation \eqref{EnergyMomentumKiselev} that in general $T^{r}_{\;\;r}\neq T^{\theta}_{\;\;\theta}$, which implies anisotropic pressure when a perfect fluid has isotropic pressure. The only case that has isotropic pressure is for the cosmological constant case with $\omega=-1$. Hence, from a conservative perspective, the Kiselev BH cannot be related to a quintessence fluid. \\

If we take $\omega_{1}=-\frac{1}{3}$, $\omega_{2}=\frac{1}{3}$ and $\omega_{3}=-1$ as nonzero state parameters in equation \eqref{LapseFKiselev}, we obtain the following lapse function

\begin{equation*}
    \Rightarrow  f(r)=1-\frac{2M}{r}-C_{[-1/3]}-\frac{C_{[1/3]}}{r^2}-C_{[-1]}r^2
\end{equation*}

Then, we can define $C_{[-1/3]}\equiv a$, $C_{[1/3]}\equiv -Q^2$, and $C_{[-1]}\equiv\frac{\Lambda}{3}$. From this, we obtain the lapse function for the dS/AdS Letelier-Reissner-Nordstrom BH, which is

\begin{equation*}
    \Rightarrow  f(r)=1-a-\frac{2M}{r}+\frac{Q^2}{r^2}-\frac{\Lambda}{3}r^2
\end{equation*}

The Kiselev BH can parameterize the most famous black holes with static spherical symmetry. Of course, this idea works with the energy-momentum tensor as well. We are not going to debate whether or not the Kiselev metric can describe the quintessence here, but we are going to exploit the ability of this metric to parameterize other black holes.

\section{BEC-KISELEV BLACK HOLE AND ITS THERMODYNAMICS}

In this section, we will generalize the discussion made in section II. This time, the graviton condensate will include matter. To do this, we extend the action given in \eqref{SmatterBEC} as follows 

\begin{equation}
    S_{BEC}=-\frac{1}{8}\int d^4x \sqrt{-\tilde{g}} (\nu(x) h^2+\mu(x) h_{\alpha \beta}h^{\alpha \beta})
\end{equation}

Where we introduced the additional term $\nu(x)h^2$. This term will allow us to get BEC-metric solutions with different matter contents. As we mentioned in section III, we do not interpret $\mu(r)$ as chemical potential anymore, so introducing another scalar field $\nu(r)$  does not affect the interpretation of our results. From this action principle, we obtain the following equation of motion 

\begin{equation}\label{EoM}
    G_{\alpha\beta}(\tilde{g}+h)= \Sigma \left(\frac{1}{2}\nu(x) h^{\sigma}_{\;\; \sigma} 
    \tilde{g}_{\alpha \beta}  + \mu(x) (h_{\alpha\beta }-h_{\alpha \sigma}h^{\sigma}_{\;\;\beta})\right)+T^{matter}_{\alpha\beta}
\end{equation}

Using equation \eqref{EnergyMomentumKiselev} as "matter" energy-momentum tensor $T^{matter}_{\alpha\beta}$, we have found the following line element

\begin{equation}\label{BECkiselevds}
    ds^2_{BEC}=-\frac{1}{(1-B)}\left(1-\frac{2M}{r}-\frac{C}{r^{3\omega +1}}\right)dt^2+\frac{1}{(1-B)\left(1-\frac{2M}{r}-\frac{C}{r^{3\omega +1}}\right)}dr^2+r^2d\Omega^2
\end{equation}

Where the scalar fields that allow us to get this solution are given by

\begin{equation} \label{scalars}
    \nu(r)=\frac{3}{4(1-B)}\frac{C\omega }{r^{3(\omega+1)}}(3\omega+1), \quad  \mu(r)=\frac{3}{2}\frac{[B(3\omega+1)-3(\omega+1)]C\omega}{(1-B)^2 r^{3+3\omega}}-\frac{1}{(1-B)^2r^2}
\end{equation}

The details of this computation are given in Appendix B. Thus, we have succeeded to generalize the BEC-Schwarzschild line element \eqref{BECdsNormal}, which describes a graviton condensate without any matter content. The line element \eqref{BECkiselevds} would describe a graviton condensate that is surrounded by different matter contents. This result is the natural extension of the metrics obtained in the reference \cite{Alfaro:2019}. We call this solution the BEC-Kiselev black hole.\\

Now, we are going to extend the result of section IV to our new graviton condensate surrounded by matter. We start computing the graviton number using equation \eqref{numberPart}, which results in

\begin{equation}\label{NBEC-Kiselev}
    N=\frac{4\pi}{3}(r_{+})^2B^2 \Rightarrow N\sim S 
\end{equation}

As before, the number of gravitons is proportional to the entropy. The radial mixed component of the effective energy-momentum tensor is given by

\begin{equation}
    T^{r}_{\;\;r}=-\frac{B}{8\pi r^2}+\frac{3C\omega(1-B)}{8\pi r^{3(\omega+1)}}
\end{equation}

Evaluating at the horizon, we obtain the thermodynamic pressure as prescribed in the HT approach

\begin{equation}\label{TotP}
    P_{tot}=-\frac{B}{8\pi r^2_{+}}+\frac{3C\omega(1-B)}{8\pi r^{3(\omega+1)}_{+}}
\end{equation}

We split the total pressure as $P_{tot}\equiv P_{vac}+P_{matt}$. Then, we have 

\begin{equation}\label{BECkiselevPmatter}
     P_{vac}\equiv-\frac{B}{8\pi r^2_{+}}, \quad  P_{matt}\equiv\frac{3C\omega(1-B)}{8\pi r^{3(\omega+1)}_{+}}
\end{equation}

The first term is the pressure related to the vacuum solution, and the second term is the pressure related to the matter content $P_{matt}$. If we take $\omega=0$, we recover the thermodynamic pressure for graviton condensate without matter given by \eqref{BECpressure} straightforwardly. We notice that apparently, if we take $\omega=-\frac{1}{3}$ and $C=1$, the total pressure given by equation (\ref{TotP}) is independent of the parameter $B$. However, this choice is not allowed because under this choice of parameters the metric has a naked singularity, explicitly, the line element becomes

\begin{equation*}
    ds^2_{BEC}=\frac{2M}{(1-B)r}dt^2-\frac{1}{\frac{2M(1-B)}{r}}dr^2+r^2d\Omega^2 
\end{equation*}

Thus, the total pressure $P_{tot}$ always depends on the parameter $B$ associated with the quantum fluctuation. Furthermore, the Hawking temperature is 

\begin{equation}\label{BECkiselevT}
    T_{h}=\frac{(1-B)}{4\pi} \left(\frac{2M}{r^2_{+}}+\frac{(3\omega+1)C}{r^{3\omega+2}_{+}}\right)
\end{equation}

We observe that the parameters $C$ and $\omega$ must be restricted in some way to assure the existence of a black hole with non-negative temperature. We give a complete discussion about the parameter space $(\omega,\lambda)$ where $\lambda\equiv\frac{C}{M^{3\omega+1}}$ in the Appendix A. It turns out that for $C\omega>0$ ($P_{matt}>0$) always there are black hole solutions, however, for $C\omega<0$ ($P_{matt}<0$) the situation is more involved. See figure 2 in Appendix A to get an insight into how the parameter space $(\omega,\lambda)$ is.

\hspace{5mm}
    
The first law of Horizon Thermodynamics can be written as

\begin{equation}
    dU=T_{h}dS-(P_{vac}+P_{mat})dV
\end{equation}

The entropy and volume have the usual definition: $S=\pi r^{2}_{+}$ and $V=\frac{4\pi r^{3}_{+}}{3}$. Besides, the internal energy is still $U=\frac{r_{+}}{2}$. These quantities have not changed with respect to the standard Kiselev BH because $r_{+}$ that marks the outer event horizon position is still located at the same place. Also, we must emphasize that we cannot use the number of gravitons $ N $ to express each thermodynamic quantity due to the presence of matter, as we did in VI. \\

We will illustrate our result for the cases of two well-known black holes. We obtain the BEC-Reissner-Nordstrom BH taking $\omega=\frac{1}{3}$ and $C_{[1/3]}=-Q^2$ in equation \eqref{BECkiselevds}. The resulting line element is

\begin{equation}\label{BECRNds}
    ds^2_{BEC}=-\frac{1}{(1-B)}\left(1-\frac{2M}{r}+\frac{Q^2}{r^{2}}\right)dt^2+\frac{1}{(1-B)\left(1-\frac{2M}{r}+\frac{Q^2}{r^{2}}\right)}dr^2+r^2d\Omega^2 
\end{equation}

For this case, the Hawking temperature is

\begin{equation}\label{BECRNT}
    T_{h}=\frac{(1-B)\sqrt{M^2-Q^2}}{2\pi(M+\sqrt{M^2-Q^2})^2}
\end{equation}

Using the second term of equation \eqref{BECkiselevPmatter}, we obtain the electrical pressure 

\begin{equation}\label{BECRNP}
    P_{matter}=-\frac{Q(1-B)}{8\pi r^{4}_{+}}
\end{equation}

The volume $ V $, the entropy $ S $, and internal energy $ U $ are functions of $r_{+}$ only; hence, they do not change with respect to the standard Reissner-Nordstrom BH. We observe that $P_{matter}$ has the same sign of $Q$. In summary, we have found the thermodynamic quantities for a graviton condensate surrounded by an electric field.\\  

Similarly, we obtain the BEC-AdS Schwarzschild black hole taking $\omega=-1$ and $C_{[-1]}=\frac{\Lambda}{3}$, which has the following line element 

\begin{equation}\label{BECAdSds}
    ds^2_{BEC}=-\frac{1}{(1-B)}\left(1-\frac{2M}{r}-\frac{\Lambda}{3}r^{2}\right)dt^2+\frac{1}{(1-B)\left(1-\frac{2M}{r}-\frac{\Lambda}{3}r^{2}\right)}dr^2+r^2d\Omega^2 
\end{equation}

The Hawking temperature and matter pressure are respectively

\begin{equation}
    T_{h}=\frac{(1-B)}{2\pi r^{2}_{+}} \left(M-\frac{\Lambda}{3} r_{+}^3\right), \quad
    P_{matter}=-\frac{(1-B)\Lambda}{8\pi}
\end{equation}

Except for the correction $(1-B)$, these quantities are precisely the same obtained in the reference \cite{Mann:2017} in the context of the extended thermodynamics approach where the negative cosmological constant plays the role of a positive pressure term. In this way, we get a graviton condensate surrounded by a cosmological fluid.

\section{Conclusions}

The N-Portrait proposal asserts that black holes physics can be understood in terms of a graviton condensate at the critical point of a quantum phase transition. Besides, it states that each thermodynamic quantity of the black hole can be expressed in terms of the number of gravitons $N$ \cite{Dvali:2011, Dvali:2012a, Dvali:2012b}. The N-Portrait analysis made by its authors is more qualitative than quantitative, further, of being not geometrical at all. On the other hand, the graviton condensate model is a geometric proposal in a more quantitative setting, which we have presented in section II. If a black hole is a graviton condensate at the critical point [$ \mu_{chem} = 0 $], it must have well-established thermodynamic variables. In this work, we have obtained a negative pressure and a correction to the Hawking temperature for the graviton condensate, which describes a black hole of parameter $ M $. Also, in the spirit of the N-Portrait proposal, we have expressed each thermodynamic variable in terms of the number of gravitons $N$. Therefore, we have successfully achieved to define the thermodynamic variables of the graviton condensate at the critical point of a quantum phase transition. \\

In the study of the line element that describes the graviton condensate, we have modified the original proposal by extending the parameter $ B $ to entire spacetime instead of being defined only inside the black hole as \cite{Alfaro:2018, Alfaro:2019}. To do this, we have accepted that spacetime is no longer asymptotically flat, but rather has a solid angle deficit of $\sqrt{1-B}\Omega$. Thanks to this, we have established a formal equivalence between the Letelier black hole with the BEC-Schwarzchild line element. In future works, we hope to exploit this equivalence more and be able to calculate the entropy of the black hole using the cloud of strings associated with Letelier's solution with mathematical tools from Loop Quantum Gravity, String Theory, or any other useful approach .\\

Another result of this work was to extend the geometrical graviton condensate model in order to include different types of matter. We have used the Kiselev black hole and its ability to parameterize other spherical solutions to pursuit this generalization. We have obtained a new solution which we called the BEC-Kiselev black hole. By choosing the pair $(\omega, C)$  appropriately in the BEC-Kiselev line element, we can obtain BEC-Reissner-Nordstrom black hole and BEC-AdS Schwarzschild black hole that describes a charged black hole and a black hole surrounded with a cosmological fluid respectively.  When there are matter contents, it cannot possibly express all the thermodynamic quantities in terms of the number of gravitons $ N $. However, this number of $N$ continues to be proportional to the entropy of the black hole. \\

One of the fundamental reasons to study a possible connection between a Bose-Einstein condensate and a black hole is the latter's own thermodynamics. The geometrical graviton condensation model of Alfaro, Espriu, and Gabbanelli was extremely successful in putting the qualitative ideas of Dvali and Gomez into geometric and quantitative terms. However, there was no thermodynamic study of the model. Furthermore, if the black hole was a condensate like any other, a valid question would be what is the volume and pressure of this condensate? Using the Horizon Thermodynamics approach, we have completed this model by obtaining its thermodynamics, including the volume and pressure terms. In addition, we have generalized the model to describe different content of matter with its respective thermodynamics. In future work, we will continue to extend the model to the quantum realm, that is, to be able to calculate the entropy of the black hole using the quantum properties of the condensate or using the equivalence between the Letelier metric and the BEC-Schwarzschild line element that describes the graviton condensate.

\begin{acknowledgements}

The work of J. Alfaro and R. Mancilla was partially supported by Fondecyt 1150390 and CONICYT-PIA-ACT1417 (Government of Chile). R. Mancilla acknowledges support from Fulbright-Chile Commission through the Fulbright BIO scholarship and from UCSB through a Fellowship for doctoral study.\\
\end{acknowledgements}

\appendix
\section{SPACE PARAMETER FOR KISELEV BLACK HOLE}

In this work, we are considering the Kiselev spacetime as a generic black hole solution in the sense that under a certain choice of the parameters $(\omega, C)$ we can recover the most well-known spherical black holes. We notice that the Kiselev solution is not the most general black hole solution under spherical symmetry, but it is enough general to obtain the most standard black hole solutions. Under this perspective, an interesting question is what is the space parameter $(\omega, C)$ that truly gives us a black hole solution. In other words, for what parameters $\omega$ and $C$ the lapse function $g^{rr}(r)=f(r)$ has a real positive root $r_{+}$. Also, we should demand black holes with positive Hawking temperature. More particularly, we should demand $f'(r_{+})>0$ where $r_{+}$ is the biggest root. 

\hspace{5mm}

We introduce the new radial variable $R\equiv\frac{r}{M}$ such that the lapse function becomes $f(R)=1-\frac{2}{R}-\frac{\lambda}{R^{3\omega +1}}$ where we have defined a new parameter $\lambda\equiv\frac{C}{M^{3\omega+1}}$. In this way, we remove the parameter $M$ and we just focus on $(\omega, \lambda)$ simplifying the analysis. We start by noticing two trivial cases where we can obtain $r_{+}$ explicitly

\begin{enumerate}
    \item For $\omega=0$, one finds that $R_{+}=2+\lambda$, then $\lambda>-2$.
    \item For $\omega=-1/3$, the root is given by $R_{+}=\frac{2}{1-\lambda}$, then $\lambda<1$.
\end{enumerate}

Now, to study the remaining cases, we see the limits of $f(R)$ as $R$ goes to infinity and zero.

\begin{itemize}
    \item In the limit $R\to \infty$, we have 2 cases:
    \begin{enumerate}
    \item For $\omega>-1/3$, we have that $f(R)\to1$ ($\forall \lambda \in \mathcal{R}$).
    \item For $\omega<-1/3$, we have that $f(R)\to-\infty sign(\lambda)$.
\end{enumerate}
    \item In the limit $R\to 0$, we have 3 cases:
    \begin{enumerate}
    \item For $0>\omega$, we have that $f(R)\to-\infty sign(\lambda)$.
    \item For $0>\omega>-1/3$, we have that $f(R)\to-\infty$ ($\forall \lambda \in \mathcal{R}$).
    \item For $-1/3>\omega$, we have that $f(R)\to-\infty$ ($\forall \lambda \in \mathcal{R}$).
    \end{enumerate}
\end{itemize}

From the study of these limits, we identify 5 regions of the space parameters $(\omega, \lambda)$ where one could have a black hole

\begin{enumerate}
    \item For $\omega>0$ and $\lambda>0$, $f(R_{+})=0$ implies $R_{+}^{3|\omega|}(R_{+}-2)=|\lambda|$. Therefore, we have one real root that satisfies $R_{+}>2$ because $f(R)$ is a continues function for $R>0$ and the limits $\lim_{R \to 0}f(R)=-\infty$, $\lim_{R \to \infty}f(R)=1$.
    \item For $\omega<-1/3$ and $\lambda<0$, $f(R_{+})=0$ implies $R_{+}^{3\omega}(2-R_{+})=\lambda$. Therefore, we have one positive real root that satisfies $0<R_{+}<2$ because $f(R)$ is a continues function for $R>0$ and the limits $\lim_{R \to 0}f(R)=-\infty$, $\lim_{R \to \infty}f(R)=\infty$.
    \item For $0>\omega>-1/3$, we have that $0<3\omega+1<0$, therefore the variable $R$ is always positive, then we have one real root because $f(R)$ is a continues function for $R>0$ and the limits $\lim_{R \to 0}f(R)=-\infty$, $\lim_{R \to \infty}f(R)=1$. 
\begin{figure}[h]
\begin{minipage}{17cm}
\includegraphics[width=12cm]{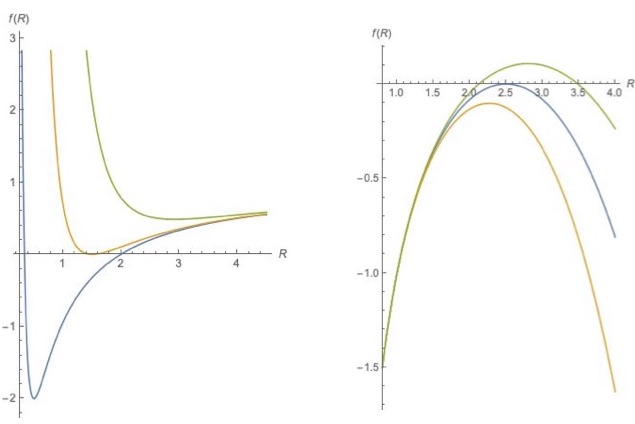}
\caption{\textbf{A}) The left graph contains three curves for $\omega>0$ and $\lambda<0$. We can see that $f''(R)$ is always positive and, for the case of two roots $R_{-}<R_{+}$, we see that $f'(R_{+})$ is always positive, therefore, the Hawking temperature is well-defined. These curves were plot with $\omega=1$, so for the extremal case $\lambda=-\frac{27}{16}$ (orange curve). \textbf{B}) The right graph contains three curves for $\omega<-1/3$ and $\lambda>0$. We can see that $f''(R)$ is always negative, then we can only have an extremal black hole (blue curve). These curves were plot with $\omega=-\frac{5}{3}$, so for the extremal case $\lambda=\frac{16}{3125}$.   }
\label{fig1}
\end{minipage}
\end{figure}

    \item For $\omega>0$ and $\lambda<0$, we have the limits $\lim_{R \to 0}f(R)=\infty$ and $\lim_{R \to \infty}f(R)=1$. In this region, one can check that $f''(R)$ is always negative, then we can have zero, one or two real roots (See Figure 1). In order to have at least one real root, we impose that $f(\bar{R})<0$ where $\bar{R}$ satisfies $f'(\bar{R})=0$. Doing this explicitly, one finds that $\bar{R}^{3\omega}=\frac{(3\omega+1)|\lambda|}{2}$. Now, imposing that $f(\bar{R})<0$, we found
    \begin{equation}
        0<-\lambda\leq\frac{2}{(3\omega+1)}\left(\frac{6\omega}{(3\omega+1)}\right)^{3\omega}
    \end{equation}
    Finally, we notice that $R_{+}>\bar{R}$. Furthermore, $f'(R_{+})>0$. Thus, the Hawking temperature is non-negative $T_{h}\geq0$. We notice that in the extremal black hole case, $T_{h}=0$ and $\lambda$ saturates the previous inequality, so
    there is only one root which is given by $R_{*}=\frac{6\omega}{(3\omega+1)}$.
    \item For $\omega<-1/3$ and $\lambda>0$, we have the limits $\lim_{R \to 0}f(R)=-\infty$ and $\lim_{R \to \infty}f(R)=-\infty$. So, again we can have zero, one or two real roots (See Figure 1), but $f''(R)$ is always positive in this region. This implies that $f'(R_{+})<0$ for the case of 2 roots $R_{-}$ and $R_{+}$ such that $R_{-}<\bar{R}<R_{+}$. However, one cannot consider this case, otherwise, the Hawking temperature would be negative. Then, in this region, one can have only extremal black holes. The relation between parameters $(\omega,\lambda)$ and the root are given by     
     \begin{equation}
        \lambda=\frac{2}{|3\omega+1|}\left(2+\frac{2}{|3\omega+1|}\right)^{3\omega}, \quad R_{*}=2+\frac{2}{|3\omega+1|}
    \end{equation}
\end{enumerate}

We resume our results in the following graphs for the parameter space $(\omega,\lambda)$:

\begin{figure}[h]
\begin{minipage}{17cm}
\includegraphics[width=17cm]{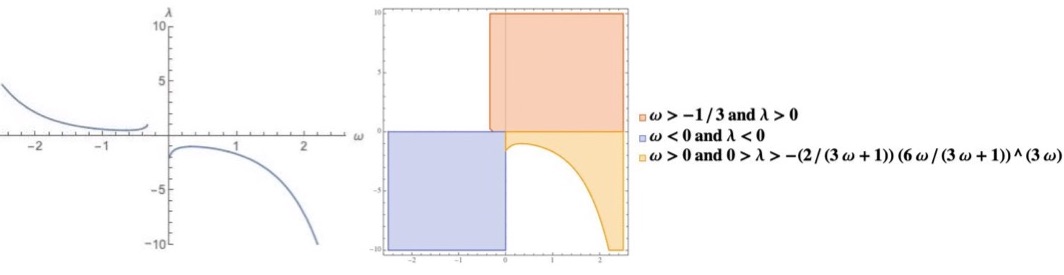}
\caption{In the graph on the left, we have 2 curves associated with the extremal black holes. In the graph on the right, we have open regions for non-extremal black holes. In the white regions, black holes solutions are not allowed because there are no real positive roots or the Hawking temperature would be negative. In addition to these 2 graphs, we have also two special case: 1) $\omega=0$, so $\lambda>-2$, and 2) $\omega=-1/3$, so $\lambda<1$.  }
\end{minipage}
\end{figure}

\section{COMPUTATION OF THE BEC-KISELEV METRIC}

We can write the most general static spherically symmetric solution as follows

\begin{equation*}
    ds^2=-f(r) dt^2+\frac{1}{h(r)}dr^2+r^2d\Omega^2
\end{equation*}

Where $d\Omega^2\equiv d\theta^2+\sin^2(\theta)d\phi^2$ is the unit 2-sphere. The nonzero mixed components of the Einstein tensor are

\begin{equation*}
    G^{t}_{\;\;t}=\frac{h-1+rh'}{r^2}
\end{equation*}

\begin{equation*}
    G^{r}_{\;\;r}=\frac{f(h-1)+rhf'}{r^2f}
\end{equation*}

\begin{equation*}
    G^{\theta}_{\;\;\theta}=G^{\phi}_{\;\;\phi}=\frac{2hff''+h'f'fr-h(f')^2r+2h' f^2+2h h'f}{4r^2}
\end{equation*}

In particular, for the Kiselev metric, we have that

\begin{equation*}
   -f(r)=-\left(1-\frac{2M}{r}-\frac{C}{r^{3\omega +1}}\right)+h_{tt} \Rightarrow f(r)=\frac{\left(1-\frac{2M}{r}-\frac{C}{r^{3\omega +1}}\right)}{1-h^{t}_{\;\;t}}
\end{equation*}

\begin{equation*}
   h(r)=\left(1-\frac{2M}{r}-\frac{C}{r^{3\omega +1}}\right)+h^{rr} \Rightarrow h(r)=(1-h^{r}_{\;\;r})\left(1-\frac{2M}{r}-\frac{C}{r^{3\omega +1}}\right)
\end{equation*}

Where we have used $h_{tt}=-f(r)h^{t}_{\;\; t}$ and $h_{rr}=\frac{h^{r}_{\;\; r}}{h(r)}$. That is to say, we are raising/lowering indices with the full metric $g_{\alpha \beta}$ and not with the background metric 
$\tilde{g}_{\alpha \beta}$. Now, we use the equation of motion (\ref{EoM}) to obtain 

\begin{equation}\label{a}
    \frac{h-1+rh'}{r^2}=\Sigma \left(\frac{1}{2}\nu(r) h^{\sigma}_{\;\; \sigma} 
    + \mu(r) h^{t}_{\;\; t} (1-h^{t}_{\;\; t} )\right)+T^{t}_{\;\;t}
\end{equation}

\begin{equation}\label{b}
    \frac{f(h-1)+rhf'}{r^2f}=\Sigma \left(\frac{1}{2}\nu(r) h^{\sigma}_{\;\; \sigma} 
    + \mu(r) h^{r}_{\;\; r} (1-h^{r}_{\;\; r} )\right)+T^{r}_{\;\;r}
\end{equation}

\begin{equation}\label{c}
  \frac{2hff''+h'f'fr-h(f')^2r+2h' f^2+2h h'f}{4r^2}=\Sigma \left(\frac{1}{2}\nu(r) h^{\sigma}_{\;\; \sigma} 
    + \mu(r) h^{\theta}_{\;\;\theta} (1-^{\theta}_{\;\;\theta} )\right)+T^{\theta}_{\;\;\theta}
\end{equation}

For the Kiselev black hole, we have that $T^{t}_{\;\;t}=T^{r}_{\;\;r}$ and $T^{\theta}_{\;\;\theta}=T^{\phi}_{\;\;\phi}$ , so by subtracting $G^{t}_{\;\;t}$ and $ G^{r}_{\;\;r}$, we obtain the following differential equation

\begin{equation*}
   f\frac{d}{dr}\left(\frac{h}{f}\right)=r \Sigma \mu(r) [h^{t}_{\;\; t} (1-h^{t}_{\;\; t} )-h^{r}_{\;\; r} (1-h^{r}_{\;\; r} )]
\end{equation*}

Then, using $f(r)$ and $h(r)$ for this case, we obtain that

\begin{equation*}
   \frac{\left(1-\frac{2M}{r}-\frac{C}{r^{3\omega +1}}\right)}{1-h^{t}_{\;\;t}}\frac{d}{dr}\left((1-h^{r}_{\;\; r})(1-h^{t}_{\;\; t})\right)=r \Sigma \mu(r) [h^{t}_{\;\; t} (1-h^{t}_{\;\; t} )-h^{r}_{\;\; r} (1-h^{r}_{\;\; r} )]
\end{equation*}

Thus, we obtain as solution for this differential equation that $h^{t}_{\;\; t}=h^{r}_{\;\; r}=constant$. Calling this constant $B$, we have that

\begin{equation*}
   f(r)=\frac{\left(1-\frac{2M}{r}-\frac{C}{r^{3\omega +1}}\right)}{1-B}, \quad h(r)=(1-B)\left(1-\frac{2M}{r}-\frac{C}{r^{3\omega +1}}\right)
\end{equation*}

Finally, we notice that equations (\ref{a}), (\ref{b}) and (\ref{c}) become an algebraic systems for $\mu(r)$ and $\nu(r)$, which explicit results are given in equation (\ref{scalars}).


\end{document}